# Multiphysics Modelling of the Molten Salt Fast Reactor using NekRS and the Fission Matrix Method

Maximiliano Dalinger, Elia Merzari, and Saya Lee

*Pennsylvania State University, Hallowell Building, State College, PA, mgd5394@psu.edu, ebm5351@psu.edu, sayalee@psu.edu*

## INTRODUCTION

The Molten Salt Reactor (MSR) was selected in 2002 by the Generation-IV International Forum as one of the reference reactor concepts due to its enhanced safety and reliability, reduced waste generation, effective fuel use, and improved economic competitiveness [1]. Since then, an innovative concept, the Molten Salt Fast Reactor (MSFR), has been proposed. This design eliminates the graphite moderator, resulting in a fast breeder reactor. The MSFR design has the particularity that the coolant is also the fuel, which tightens the coupling between neutronics and thermal hydraulics as the fuel circulates through the primary system. Therefore, developing computational models to analyze the MSFR requires a multiphysics approach. In this paper, we propose developing a neutronic–thermal-hydraulic computational model of the MSFR that uses a reduced-order model to solve the neutronics equations. The principal computational tool chosen for this purpose is the high-fidelity code Cardinal, a wrapping within the MOOSE framework that integrates the Computational Fluid Dynamics code NekRS and the Monte Carlo particle transport code OpenMC. However, we use the Fission Matrix (FM) Method to solve the neutronics equations instead of OpenMC.

## METHODOLOGY

### NekRS

NekRS [2] is a GPU-accelerated open-source Spectral Element Method (SEM) Computational Fluid Dynamics (CFD) code to simulate fluid dynamics and heat transfer. It solves the unsteady incompressible Navier-Stokes and energy equations in 2-D or 3-D geometries. Its capabilities include solving Conjugated Heat Transfer (CHT) problems, Reynolds-Averaged Navier Stokes (RANS) modeling, Large Eddy Simulation (LES), and Direct Numerical Simulation (DNS). It supports both CPU and GPU backends. In this work, we use the RANS $k-\tau$ turbulence model.

### MOOSE

MOOSE [3] (Multiphysics Object-Oriented Simulation Environment) is a finite element framework for solving fully coupled, fully implicit multiphysics simulations. It executes multiple sub-applications simultaneously and transfers data between the scales. It discretizes the space using the PETSc non-linear solver and libMesh. Some of its capabilities include automatic differentiation, scaling to a large number of processors, hybrid parallelism, and mesh adaptivity.

### Cardinal

Cardinal [4] is an open-source application that wraps OpenMC and NekRS codes within the MOOSE framework. It uses the MOOSE data transfer implementation to perform high-fidelity coupled neutronic-thermal hydraulics calculations. It can also be coupled with any MOOSE-based application, enabling broad multiphysics capabilities.

### The Fission Matrix Method

A fission matrix is an *n x n* matrix *A* which corresponds to a discretization of the reactor geometry into *n* spatial cells. Each element of this matrix, $a_{ij}$, represents the number of neutrons generated in cell *i* from a fission neutron born in cell *j*. The fundamental eigenvector of *A* is the fission source distribution $\vec{F}$, and the fundamental eigenvalue is the multiplication factor $k_{eff}$. The FM Method formulation [5] is shown in Equations 1.1 and 1.2. Solving the FM means obtaining the $k_{eff}$ and the fission source distribution.

$$\vec{F} = \frac{1}{k_{eff}} A \vec{F} \qquad (1.1)$$

$$F_i = \frac{1}{k_{eff}} \sum_{j=1}^{N} a_{ij} F_j \qquad (1.2)$$

## COMPUTATIONAL MODEL

The model used in this project is based on reference [6], which couples the NekRS solution with OpenMC using Cardinal. Here, the difference is that the FM method is used to solve the neutronics problem rather than OpenMC.

### NekRS MSFR Model

NekRS solves fluid mass, momentum, and energy conservation using a RANS $k-\tau$ turbulence model. It receives the volumetric heat source calculated in OpenMC. Figure 1 shows the three-dimensional geometry of the MSFR primary loop used in the calculations, generated with Gmsh. The model uses simplified geometry with a continuous inlet/outlet and does not model the pumps or heat exchangers.

The separation between the inlet and the outlet is 50 cm. The boundary conditions for the energy equation are a uniform inlet temperature of 898 K and adiabatic walls; the code calculates the outlet temperature. Regarding the momentum equation, the boundary conditions are a parabolic velocity profile with a mean non-dimensional value of 1 at the inlet, non-slip at the remaining walls, and the code calculates the outlet velocity. For turbulence variables $k$ and $\tau$, parabolic profiles were used at the inlet with mean non-dimensional values of 0.01 and 0.1, respectively. Table I lists the relevant non-dimensional parameters used in the simulations. The mesh has $1.35 * 10^6$ hexahedral elements. A polynomial order 3 was selected, resulting in about $8.6 * 10^7$ Gauss-Lobatto-Legendre (GLL) quadrature points. This resolution was found adequate for the problem. A 2D RANS version of the NekRS model was verified against OpenFOAM in [7], showing a good agreement for the velocity field.

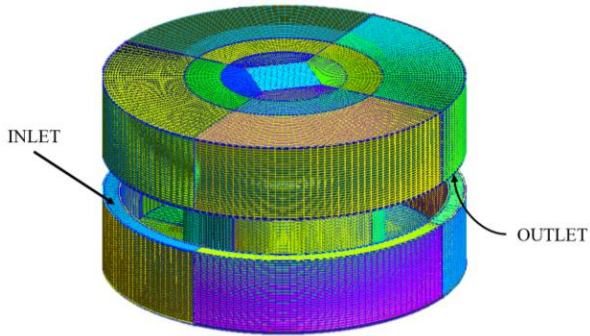

Fig. 1. NekRS mesh of the MSFR.

Table I. Nondimensional parameters used for NekRS coupled calculations.

| Nondimensional Parameter | Symbol | Value |
|---|---|---|
| Reynolds Number | $Re$ | 30672 |
| Prandtl Number | $Pr$ | 16 |
| Peclet Number | $Pe$ | 491286 |
| Turbulent Prandtl Number | $Pr_t$ | 1 |

**FM Databases**

FMs are obtained from Monte Carlo simulations for a specific reactor state, i.e., a particular fuel temperature distribution. Therefore, to accurately represent the reactor state, which differs from that used in Monte Carlo simulations, multiple FMs are required, called Fission Matrix Databases (FMDBs). The FM representative of the reactor's actual state is derived from the FMDBs.

The Monte Carlo code Serpent was used to generate the FMDBs. The 3D model of the MSFR is shown in Figure 2. The core is a cylinder surrounded radially by a fertile blanket (blue) of LiF-ThF$_4$ (77.5-22.5%mol), and a boron-carbide layer (green). All of it is surrounded by a stainless-steel reflector (gray). The salt (orange) is composed of LiF-ThF$_4$-$^{233}$UF$_3$ (77.5-19.65-2.85%mol). In all cases, Serpent calculations used 50000 particles, with 1000 active cycles and 20 inactive cycles, resulting in uncertainties below 7.1 pcm for Keff and a maximum RMS of 0.9% for the fission source distribution. The cross-section library used was ENDF/B-VII.1. The Serpent model was verified against OpenMC in [6]. FMDBs were calculated in a 3D geometry with a 13x13x13 Cartesian discretization using Serpent. Five FMDBs were generated using uniform temperature profiles for 800 K, 900 K, 1000 K, 1100 K, and 1200 K.

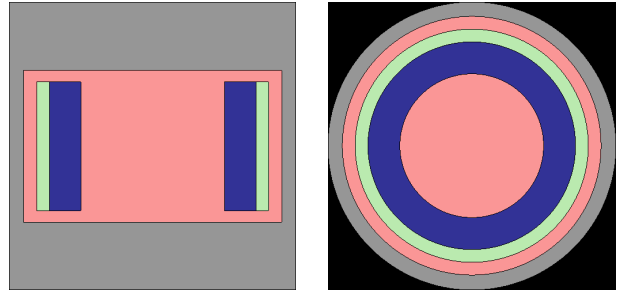

Fig. 2. Serpent model of the MSFR. Lateral view (left) and superior view (right).

**FM Linear Interpolation**

Matrix $A$ is calculated for a particular fuel temperature. To do this, each element $a_{ij}$ is linearly interpolated between the two nearest FMDBs. Interpolations follow Equation 2, where $d$ and $d+1$ are the nearest database indexes, and $T$ is the fuel temperature. This procedure has been previously analyzed [8, 9].

$$a_{ij} = \tilde{a}_{ij}^{(d+1)}\left(\frac{T_i - T^{(d)}}{T^{(d+1)} - T^{(d)}}\right) + \tilde{a}_{ij}^{(d)}\left(1 - \frac{T_i - T^{(d)}}{T^{(d+1)} - T^{(d)}}\right) \quad (2)$$

**FM Model**

To use the FM method in MOOSE, some code modifications were required. Functions were developed in C++ and introduced in MOOSE to load the FMDBs, perform the FM linear interpolation, solve the FM, and return the fission source distribution and multiplication factor. The FM model was developed in MOOSE. It consisted of a Cartesian mesh of 13×13×13 elements, similar to the Cartesian grid used to generate the FMDBs. A temperature distribution must be provided so that each cell receives a constant temperature. The FM for that temperature distribution is obtained by linear interpolation, then solved to obtain the multiplication factor and the fission source distribution. The volumetric heat source distribution is then calculated by scaling the normalized fission source to the desired total power. Figure 3 shows how the NekRS mesh fits inside the FM mesh. As shown, some meshes will receive no feedback from NekRS. Additionally, since there is no fuel material in these regions, the heat source will be null in those cells. The total power was 3e8 W.

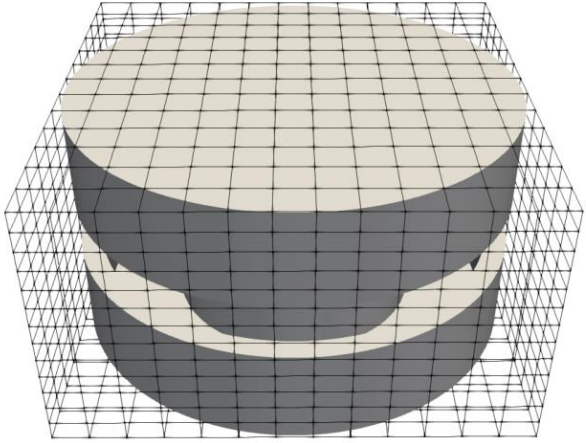

Fig. 3. NekRS mesh fitting inside the FM mesh.

**Cardinal multiphysics Coupling**

Figure 4 presents the coupling procedure used in Cardinal, often called the Picard iteration "in time" [4]. The FM model runs first using an initial uniform temperature and density. It calculates the fission source distribution, and it is rescaled to obtain the heat source $q'''_{fis}$. Cardinal receives this information and sends it to NekRS. Then NekRS performs $N$ calculations of the time step $\Delta t_{nek}$ to calculate fluid temperature and velocity. NekRS sends the fluid temperature to Cardinal. Temperatures are updated in the FM model, and the process is repeated until convergence. Convergence is reached when the maximum, average, and outlet temperatures reach steady-state values. For the present work, $N = 10000$ and $\Delta t_{nek} = 2.0 * 10^{-4}$. Therefore, the FM model receives feedback only for the fuel temperature.

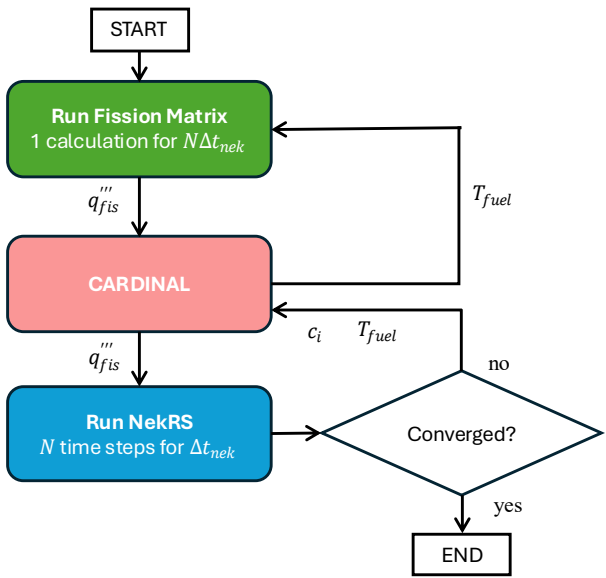

Fig. 4. Cardinal coupling scheme.

## RESULTS

Simulations in Cardinal were run on the supercomputer Frontier, using GPUs for NekRS and CPUs for the FM model. The results presented here are preliminary. The multiplication factor obtained with the FM model is 1.04632. The simulation was run for approximately 60 convective units, and then time averaging was performed for approximately 80 more convective units.

Figure 5 presents the instantaneous heat-source distribution from the last time step in the FM model, and the one used in NekRS. We can see that it has a sinusoidal shape. Figure 6 presents the time-averaged velocity distribution obtained in NekRS. This is useful for identifying stagnation regions in the periphery of the central cylinder, where the temperature reaches its maximum. Figure 7 presents the time-averaged temperature distribution obtained in NekRS and used in the FM model. We can see that the temperature rises in stagnation regions because the flow is "trapped" and continues to be heated. The outlet temperature obtained in NekRS is 997.4 K, and the average temperature is 973.0 K.

In reference [6], using a NekRS-OpenMC model, the outlet temperature is 997.3K, and the average temperature is 983.1 K. This means that the NekRS-FM model differs by 0.1 K and 10.1 K, respectively, from the NekRS-OpenMC model. This small difference is expected since the total energy is the same in both models. Comparison with reference temperature, velocity, and heat-source distributions shows good agreement.

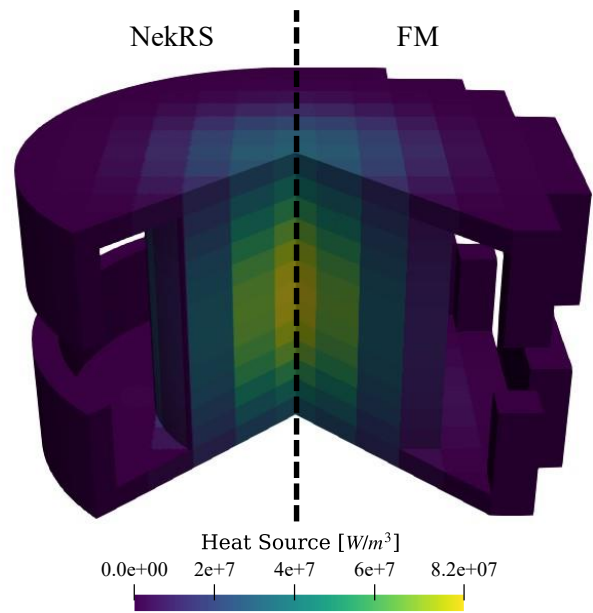

Fig. 5. Instantaneous heat source obtained with the FM model (right) and used in NekRS (left).

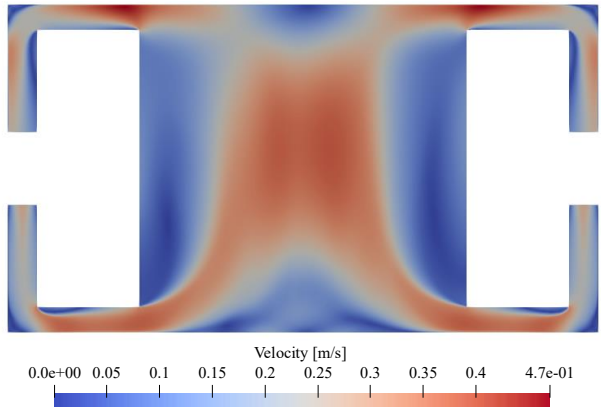

Fig. 6. Average velocity distribution obtained in NekRS.

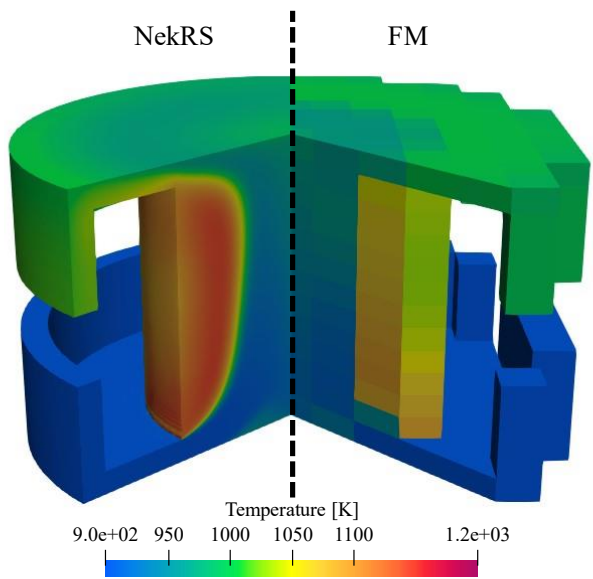

Fig. 7. Average temperature obtained in NekRS (left) and used in the FM model (right).

## CONCLUSIONS

A multiphysics model of the MSFR has been developed in Cardinal, accounting for neutronic–thermal-hydraulic feedback. Neutronic equations were solved with the Fission Matrix method using a reduced order model, while mass and momentum equations were solved in NekRS. Preliminary results showed reasonable behavior for the heat source, temperature, and velocity. Low-velocity regions act as stagnation zones, where the temperature reaches its maximum. Comparisons with a previous NekRS-OpenMC model showed good agreement for temperature, velocity, and heat-source distributions.

Further analysis will include a sensitivity analysis for the mesh used in the FM model, testing coarser and finer meshes to identify the optimal discretization. Additionally, we want to analyze the influence of the mesh shape on the results using a cylindrical mesh.

## ACKNOWLEDGMENTS

This research was in part performed using funding received from the U.S. NRC, Grant #31310022M0037.

## REFERENCES

1. P. RUBIOLO, et. al., "Preliminary Thermal-Hydraulic Core Design of the Molten Salt Fast Reactor (MSFR)," *Annals of Nuclear Energy*, **64**, 449 (2014).
2. P. FISCHER, et. al., "NekRS, a GPU-accelerated Spectral Element Navier–Stokes Solver," *Parallel Computing*, **114**, 102982 (2022).
3. C. J. PERMANN, et. al., "MOOSE: Enabling massively parallel multiphysics simulation," *SoftwareX*, **11**, 100430 (2020).
4. A. NOVAK, et. al., "Coupled Monte Carlo and Thermal-Fluid Modeling of High Temperature Gas Reactors Using Cardinal," *Annals of Nuclear Energy*, **177**, 109310 (2022).
5. S. CARNEY, F. BROWN, B. KIERDROWSKI, and W. MARTIN, "Theory and applications of the fission matrix method for continuous-energy Monte Carlo," *Annals of Nuclear Energy*, **73**, 423, (2014).
6. M. DALINGER, et. al., "High-Fidelity Modelling of the Molten Salt Fast Reactor", *Proceedings of the 21st International Topical Meeting on Nuclear Reactor Thermal Hydraulics (NURETH-21)*, Korea (2025).
7. J. ACIERNO, et. al., "Development and Application of Reduced-Order Models for Thermal-Fluid Dynamics in Molten Salt Reactors," *Nuclear Technology*, (2025).
8. A. RAU and W. WALTERS, "Validation of coupled fission matrix – TRACE methods for thermal-hydraulic and control feedback on the Penn State Breazeale Reactor", *Progress in Nuclear Energy*, **123**, 103273, (2020).
9. T. TOPHAM, A. RAU, and W. WALTERS, "An iterative fission matrix scheme for calculating steady-state power and critical control rod position in a TRIGA reactor," *Annals of Nuclear Energy*, **135**, 106984, (2020).